\documentclass[twocolumn,showpacs,preprintnumbers,amsmath,amssymb]{revtex4}

\usepackage{graphicx}
\usepackage{dcolumn}
\usepackage{bm}

\DeclareMathOperator{\arcsec}{^{\prime\prime}}
\DeclareMathOperator{\arcmin}{^{\prime}}

\begin{document}

\preprint{arXiv:0803.0027v2 [astro-ph]}

\title{Search for Cosmic Strings in the GOODS Survey}

\author{J. L. Christiansen}
\email{jlchrist@calpoly.edu}
\affiliation{Department of Physics, California Polytechnic State University, 
San Luis Obispo, California 93407, USA}
\author{J. Goldman}
\affiliation{Department of Physics, Faculty of Science, 
National University of Singapore, Singapore, 117542}
\author{E. Albin}
\affiliation{Department of Physics, California Polytechnic State University, 
San Luis Obispo, California 93407, USA}
\author{K. A. James}
\affiliation{Department of Physics, California Polytechnic State University, 
San Luis Obispo, California 93407, USA}
\author{D. Maruyama}
\affiliation{Department of Physics, University of California, 
Berkeley, California 94720, USA}
\author{G. F. Smoot}
\affiliation{Lawrence Berkeley National Laboratory, Space Sciences Laboratory 
\\and Department of Physics, University of California, Berkeley, California 94720, USA}



\date{\today}

\begin{abstract}
We search Hubble Space Telescope Treasury Program 
images collected as part of  
the Great Observatories Origins Deep Survey 
for pairs of galaxies consistent with the gravitational lensing signature 
of a cosmic string.  
Our technique includes estimates of the efficiency for finding the 
lensed galaxy pair.
In the north (south) survey field we find no evidence out to a redshift of 
greater than 0.5 (0.3) for cosmic strings to a mass per unit length limit of
$G\mu/c^2<3.0\times10^{-7}$ at 95\% confidence limits (C.L.).
In the combined 314.9 arcmin$^2$ of the north and south survey fields 
this corresponds to a global limit on $\Omega_{strings}<0.02$.  
Our limit on $G\mu/c^2$ is more than an order of magnitude lower than 
searches for individual strings in cosmic microwave background (CMB) data.  
Our limit is higher 
than other CMB and gravitational wave searches, however, we note that it 
is less model dependent than these other searches.

\end{abstract}


\maketitle

\section{\label{sec:intro}Introduction:}\protect
Cosmic strings can be formed during symmetry-breaking phase transitions in the 
early universe \cite{kibble-GUT, polchinski-DF}.
Their well defined equations of motion and interaction potentials give us 
reason to believe that they have evolved into a modernday string network, 
observable through a variety of astrophysical phenomena \cite{hindmarsh}. 
Proposed as a natural consequence of the cooling universe, they were 
originally believed to be an unavoidable byproduct of symmetry breaking at 
the grand unified theory (GUT) scale and it was thought that measurement 
of their mass per unit length would tell 
us the temperature of phase transitions. 
More recently, cosmic (super) strings have been proposed as a byproduct of 
supersymmetric F- or D-term inflation, occurring after the GUT scale 
transition and resulting in a stochastic network with interaction 
probabilities less than 1 due to extra dimensions that allow strings to pass 
without touching \cite{sakellariadou}.
In either case, the dimensionless scale of observational interest is from 
$10^{-6} \gtrsim G\mu/c^2 \gtrsim 10^{-11}$.  

Although there has been considerable interest within the theory community, 
only a few observations bear on the subject: 
(1) cosmic microwave background (CMB), (2) gravitational waves, 
(3) gravitational lensing.
The CMB power spectrum shows that cosmic strings are not the dominant 
factor in large-scale structure formation,
contributing less than 10\% of the observed structure \cite{WMAP-1, WMAP-2, WMAP-3, WMAP-4}.  
Searches for individual strings in the CMB set a limit 
$G\mu/c^2 \lesssim 3.7\times10^{-6}$ \cite{jeong, lo}. 
Bursts of gravitational waves are predicted from cusps in
cosmic strings as they acquire a large Lorentz boost due to the string tension.  
A population of cusps and loops is expected to produce a stochastic background 
of gravitational waves that can be detected
via pulsar timing and also by direct measurement with LIGO 
\cite{pulsar, ligo}.  
The lack of a gravitational wave signal sets a limit on cosmic strings 
masses, $G\mu/c^2 \lesssim 1.5\times10^{-8}$.  
This limit depends on properties of the string network such as the physical 
model, number, and strength of interactions.
Gravitational lensing by a cosmic string of background galaxies has also been 
considered.    
A candidate pair of morphologically similar galaxies, CSL-1, was discovered 
in March of 2002 \cite{csl-1, csl-2} 
but follow-up Hubble images proved it to be a binary system 
\cite{csl-1-refuted}.
A systematic search of an optical survey field as we report here has not previously 
been published.

Our aim in this paper is to use the  
wide-and-deep-field survey carried out by the 
Great Observatories Origins Deep Survey (GOODS) team 
with the Hubble Space Telescope (HST) Advanced Camera for Surveys (ACS) 
to search for the existence of cosmic strings using their lensing signature. 
We have developed the observational technique beyond what has been previously 
attempted \cite{csl-1, oguri} by including the efficiencies of finding 
the lensed galaxy in our analysis.  
We note that the string masses excluded with this technique are lower than 
those ruled out by the CMB search for individual strings.  
Still, the CMB full-sky observations at very large redshifts search a larger 
volume of the universe.
Searching for less massive strings in high resolution wide-field surveys is 
important, however, because in many models the density of strings increases 
logarithmically with decreasing mass.  
Recently reported sensitivities to cosmic strings suggest that modern optical
surveys are competitive with other methods \cite{gasparini}.
At the present time, limiting the mass scale and density of cosmic strings is 
important for model development.  Ultimately, the 
discovery of cosmic strings would be an important key to the 
physics of the early universe.

In Sec. II we give an overview of the lensing technique.  
Section III follows with a description of our data selection.
Section IV then describes the simulation of galaxy lensing by cosmic strings 
that is used in Sec. V to determine our detection efficiencies.  
The efficiencies are used in Sec. VI to determine limits on individual 
cosmic strings as a function of mass and redshift as well as the limit on 
the density of cosmic strings in the universe.  
We summarize our results in Sec. VII.

\section{\label{sec:overview}Overview of Lensing Technique:}\protect
The geometry of space is altered by the large mass per unit length 
of a cosmic string.  A long straight cosmic string will cause an angular
defect or deficit according to
\begin{equation}\label{eq:geo}
ds^2 = dz^2 + dr^2 + (1-4\frac{G\mu}{c^2})^2r^2d\theta^2
\end{equation}
where the coordinate $z$ is along the string, and $r$ and $\theta$ are the 
polar coordinates of a plane perpendicular to the string \cite{peebles}.  
The deficit angle is given by the dimensionless parameter, 
$\delta = 8\pi\frac{G\mu}{c^2}$ which results in the lensing effect on 
background galaxies, making identical pairs appear on both sides of the string.
The opening angle between the two observed images is related 
to the deficit angle by
\begin{equation}\label{eq:oa}
\Delta\theta = \delta sin(\beta)\frac{D_{ls}}{D_{os}}
\end{equation}
where $D_{ls}$ is the distance between the lensing string (l), and the 
background source (s), $D_{os}$ is the distance between the observer (o) 
and the background source, and $\beta$ is the tilt of the string 
toward the observer \cite{ryden}.

\begin{figure}
\includegraphics{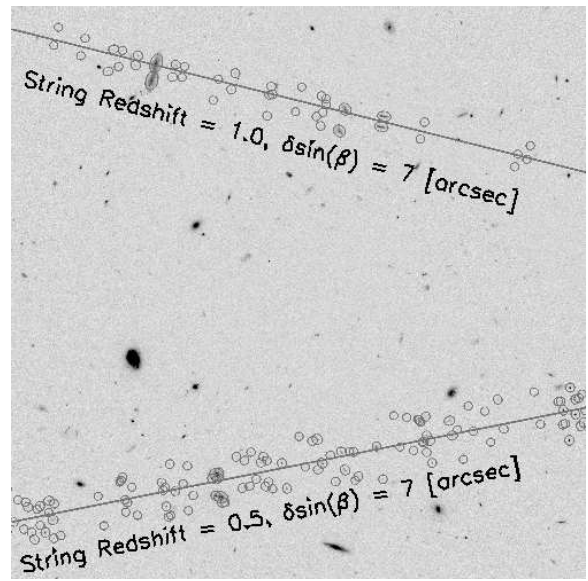}
\caption{\label{fig:overviewString} Simulated cosmic strings at redshift 0.5 
and 1.0 in a small part of the GOODS north field. 
Pairs of morphologically similar galaxies are expected to fall on 
opposite sides of the string.}
\end{figure}

\begin{figure}
\includegraphics{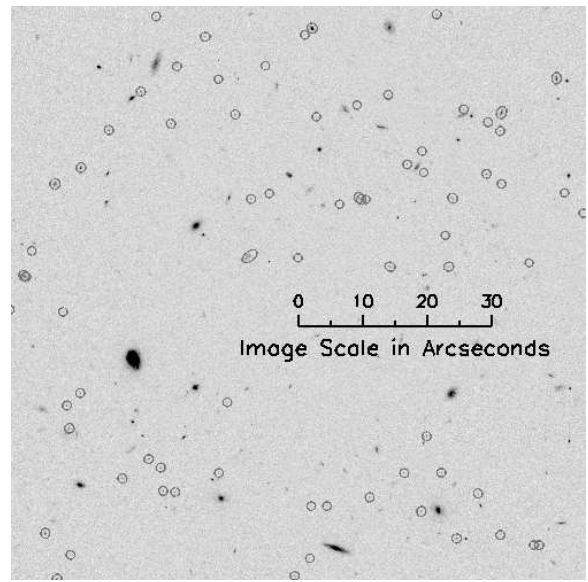}
\caption{\label{fig:overviewSearch}Pairs of morphologically similar galaxies
found in a region of the GOODS survey.  These pairs form the background to 
the signal pairs from a cosmic string.}
\end{figure}

Our strategy is to search the GOODS 
wide-field survey for all pairs of galaxies that are morphologically 
similar -- the hallmark of a cosmic string -- with opening angles less 
than $15\arcsec$.  
Figure \ref{fig:overviewString} shows simulated pairs of galaxies produced by 
massive cosmic strings at redshifts of 0.5 and 1.0 in a small part of the 
survey.  
In contrast, Fig. \ref{fig:overviewSearch} shows the random pairs 
of morphologically similar galaxy pairs that comprise the background to the 
cosmic string search.  


The difference between the signal and background is characterized statistically
by the number of morphologically similar galaxy pairs as a function of the 
opening angle.  
Pairs associated with a string are expected to pile up at angles, 
$\Delta\theta$, less than about $6\arcsec$.
The distribution depends on mass, redshift, and tilt of the string as well as 
specifics of the survey which are further discussed in Sec. \ref{sec:sim}.  
The background consists of a small number of galaxies that happen to be 
morphologically similar. 
The number of random pairs is approximately proportional to the area of 
an annulus, $d\Omega$, defined by a bin extending from $\Delta\theta$ to 
$\Delta\theta + \delta\theta$.  
\begin{equation}\label{eq:annulus}
d\Omega = \pi((\Delta\theta + \delta\theta)^2 - \Delta\theta^2)
\end{equation}
which, for small bin sizes reduces to a very nearly linear rise with 
opening angle, $d\Omega = 2\pi\delta\theta\Delta\theta$.

\section{\label{sec:data}Data Sample:}\protect

We analyze the GOODS Version 1.0 
fits images taken with the ACS aboard HST \cite{GOODS}. 
Two fields are available, the Hubble Deep Field North (HDF-N) and the Chandra 
Deep Field South (CDF-S) which we refer to as the north and south fields 
respectively.  

\subsection{Source identification}
SExtractor (Source Extractor) has, over the past several years, become the 
standard tool used in the analysis of space telescope data to identify 
sources.
It has evolved from the ``quick and dirty'' solution envisioned by its 
author\cite{SEX-1} to a program used worldwide by those conducting 
cosmological and astronomical research. 
We use the catalogs produced by an unmodified copy of SExtractor v2.5.0 in 
the analysis described here \cite{SEX}.

SExtractor's function is two-fold; namely, it processes FITS image files 
both to discover sources and, at the same time, to perform 
photometric calculations based on the data from those sources. 
Essentially, it locates stars, galaxies, and nebulae  and 
makes estimates of their size, shape, brightness, and surrounding background 
exposure levels. 

Calculating photometric quantities requires adjustment of a broad range of 
parameters specified in the SExtractor configuration file. For the GOODS 
I-band (F775W) north and south fields (the focus of this analysis), 
a specific set were chosen that maximize the efficiency of discovery while 
rejecting as many spurious objects as possible. 
A brief description of the relevant parameters and their values follows.

The \verb+ANALYSIS_THRESH+ and \verb+DETECT_THRESH+ parameters specify the level 
above background at which we set the pixel thresholds for the photometric/analytical 
and detection algorithms, respectively. 
For the analysis of the GOODS data, these two parameters are set to 
1.0 times the RMS of the background level. 
The deblending threshold parameter, \verb+DEBLEND_NTHRESH+, determines how 
aggressively SExtractor is in its attempts to subdivide an agglomeration of 
above-background 
pixels into subgroups corresponding to smaller, closely-spaced objects. 
The quantity 
\verb+DEBLEND_NTHRESH+ itself represents the number of brightness thresholds used in 
this procedure. 
We have used the (default) value of 32. 
The associated parameter 
\verb+DEBLEND_MINCONT+ determines how bright a particular group of 
pixels must be to qualify as an independent object with the value 0 causing maximal 
deblending and the value 1 causing no deblending at all. 
For this analysis, we have set \verb+DEBLEND_MINCONT+ to 0.01. 
The minimum area a grouping of pixels must have in order 
to be considered an object is specified by the \verb+DETECT_MINAREA+ parameter. 
In this analysis this minimum area is 9 pixels.

The process of weighting in SExtractor is flexible, and many options exist. 
For our analysis we have found that the \verb+MAP_WEIGHT+ option, 
which requires a weight image 
accompanied by \verb+WEIGHT_THRESH+ of 0000000,0000000, 
gives the best results. 
Images, specified with \verb+WEIGHT_IMAGE+ on the SExtractor command line, 
are available alongside the scientific image files on the GOODS website. 
Apart from the parameter settings outlined here, all other inputs available 
in the SExtractor parameter file are set to their default values. These 
values are defined and discussed in detail in the SExtractor 
manual \cite{SEX-1}.

\subsection{\label{sec:dataGal}Galaxy selection:}\protect
The catalog created by SExtractor for the I-band fits images 
contains 51 538 objects in the north field and 45 208 objects in the 
south field.
To eliminate identification of spurious objects, we only select 
galaxies within a rectangular fiducial region where the exposure
time is relatively uniform.
The total area analyzed is 159.5 arcmin$^2$ in the north field 
and 155.4 arcmin$^2$ in the south field.
We also remove stars from our own galaxy by requiring CLASS\_STAR $ < 0.9$. 

We further post-process our galaxy catalog by applying a procedure 
to identify the 
pixels in the image associated with each galaxy.  
The first step is to define a small but encompassing search region about each 
galaxy centroid and to smooth the region so that we are less sensitive to noise.
We use a standard Lee filter in IDL for this (LEEFILT \cite{LEEFILT}).
The second step is to find a bright pixel near the galaxy centroid.  
Then we look for neighboring pixels that are $1 \sigma$ above the noise 
threshold in the unsmoothed image and attach them to the centroid pixel cluster.  
By iteratively connecting neighbors that are above the noise threshold, we 
eventually get a cluster of pixels that we identify as the galaxy.  
This process sometimes merges neighboring galaxies.  
In the event that a cluster of pixels reaches the edge of the search region 
or that two galaxies merge, we raise the neighbor threshold to $2 \sigma$ and 
repeat the process until each galaxy is completely contained within the search
region and does not contain the centroid from any other galaxy in the catalog.
For a few very dim sources, the threshold is raised so high that there are no 
pixels left in the cluster and we remove these galaxies from the sample.  
After fiducial cuts, star removal, and pixel ID, the 
resulting catalogs contain 41 358 galaxies in the north and 36 328 galaxies 
in the south.

\subsection{\label{sec:dataCorr}Galaxy-galaxy correlation:}
\begin{figure}
\includegraphics{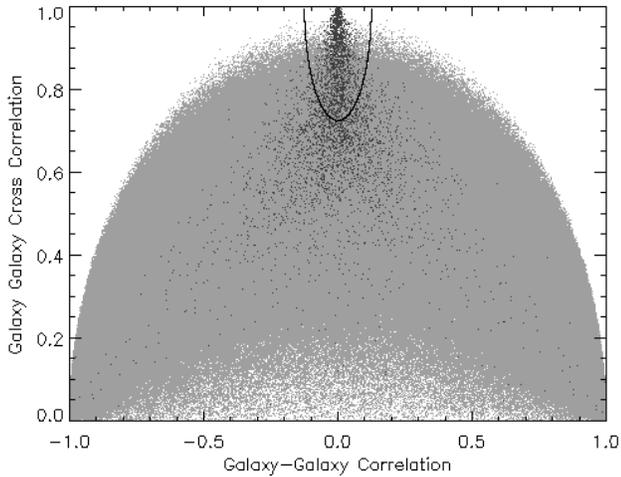}
\caption{\label{fig:corr} 
Correlations for random pairs of galaxies in the GOODS data (light gray points)
compared to simulated signal (black points).  
We define matched galaxy pairs as those within the elliptical line at the
top of the plot where the correlation and cross-correlation are optimal.}
\end{figure}

\protect
To calculate the morphological similarity between each pair of galaxies we 
rely on the correlation and cross-correlation of the two galaxy images.  
This is a reasonably optimal way to assess the similarity of both brightness
and shape.
We first align the centroids and then calculate the correlation ($CORR$) and 
the cross-correlation ($XCORR$) of the pixels.
\begin{equation}\label{eq:corr}
CORR = \frac{\sum{I_1(x_i,y_i)^2} - \sum{I_2(x_i,y_i)^2}}{\sum{I_1(x_i,y_i)^2} + \sum{I_2(x_i,y_i)^2}}
\end{equation}
\begin{equation}\label{eq:xcorr}
 XCORR = \frac{2\sum{I_1(x_i,y_i)*I_2(x_i,y_i)}}{\sum{I_1(x_i,y_i)^2} + \sum{I_2(x_i,y_i)^2}}  
\end{equation}
where $I(x_i,y_i)$ is the intensity of each pixel in a galaxy and 
the subscript 1 or 2 refers to the galaxies being correlated.
Perfectly correlated galaxies will have identical intensity distributions so that 
$CORR = 0$ and $XCORR = 1$.
Measurement noise will smear the distributions out somewhat.  
Figure \ref{fig:corr} shows the XCORR vs CORR distributions for both signal
and background.  
The signal is concentrated as expected at the top and center but also has 
a long broad tail downward due to dim galaxies that are especially 
sensitive to noise.
We have also found that near our detection threshold, galaxies tend to contain very
few pixels and become round in appearance regardless of their true shape.  
The background is concentrated at $CORR \sim \pm 1$ 
and $XCORR \sim 0$, however, the statistical nature of the distribution extends
between these limits in a semicircular pattern.

\subsection{\label{sec:dataMatch}Matched galaxy pair selection:}\protect
We define matched galaxy pairs as those within the elliptical line drawn in 
Figure \ref{fig:corr}.  
This cut defines our definition of morphologically
similar galaxies and was designed to maximize our signal pairs relative to 
background pairs.  
Although we would like to be more efficient for signals, we find that the
signal outside this cut is swamped by background.

In this analysis, we consider pairs of galaxies with opening angles, 
$\Delta\theta<15\arcsec$. 
There are 3 668 matched pairs in the north field and 2 978 matched pairs 
in the south field with $\Delta\theta<15\arcsec$.  
This is compared to 5 091 501 total pairings in the north field and 4 180 440 
total pairings in the south field with $\Delta\theta<15\arcsec$.  
The correlation cuts therefore reduce the background by a factor of 1 400.
A small region of the north field is shown in Fig. \ref{fig:overviewSearch}. 
The figure indicates that pairs passing our cuts appear fairly randomly 
distributed, consistent with random pairings that constitute our background.

\subsection{\label{sec:pairs}Pairs distribution:}\protect

\begin{figure}
\includegraphics{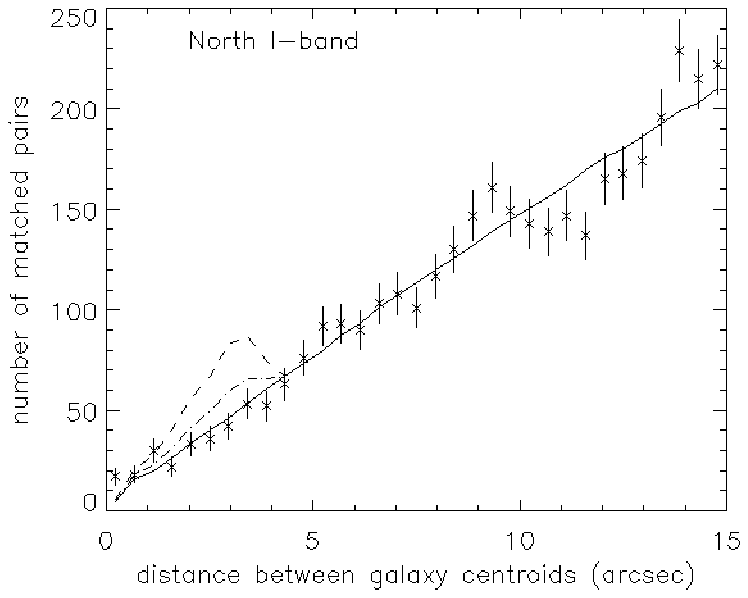}
\includegraphics{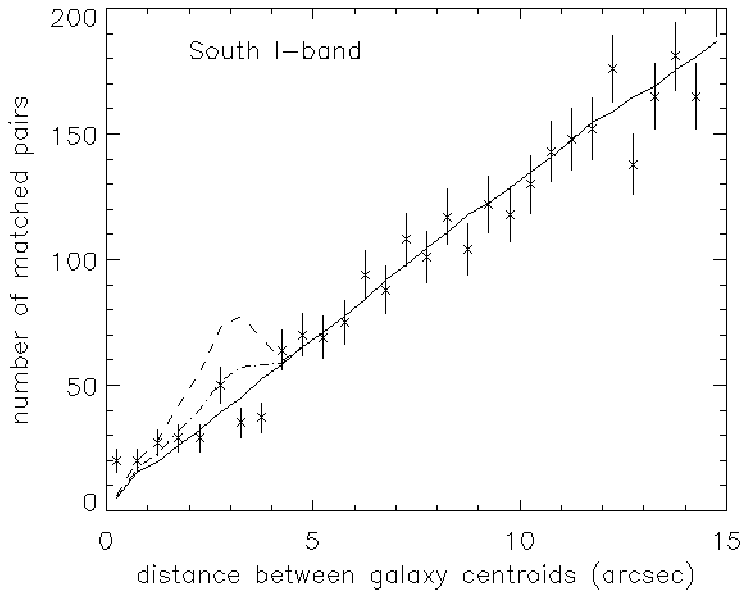}
\caption{\label{fig:pairs}
Pairs of galaxies in the GOODS north- and south-field data (points) 
compared to background (solid line) and one example of a simulated 
string (dashes).  
The upper simulated string is the total number of pairs expected from the 
simulation with string redshift of 1.2 and $\delta\sin\beta$ of $6\arcsec$.  
The lower simulated string includes measurement inefficiencies.}
\end{figure}

The binned distribution of matched galaxy pairs is shown in 
Fig. \ref{fig:pairs}. 
The background shape is characterized by the distribution of all pairs 
of galaxies 
regardless of size and shape.  
Because strings with masses large enough to create opening angles greater 
than $7\arcsec$ have been ruled out \cite{WMAP-1,jeong}, 
we normalize the background distribution 
to the number of measured matched pairs between  $7\arcsec$ and $15\arcsec$. 
This gives us a reliable estimate of the background at smaller 
opening angles.  
From the background, we observe that SExtractor merges galaxies with opening
angles smaller than $0.4\arcsec$.  At slightly larger opening angles there 
is a tiny excess of pairs created when SExtractor splits lumpy sources 
into two sources.  
We include this effect in the analysis, but note that the number of excess 
matched pairs is negligible.
In our signal region, between $0.4\arcsec$ and $7\arcsec$, 
the $\chi^2/dof$ of the matched pairs to the background is 
1.3 for the north field and 1.1 for the south field.
Based on the scaled background distribution, we see no evidence for an
excess of pairs at small opening angles.

\section{\label{sec:sim}Signal Simulation:}
\begin{figure}
\includegraphics{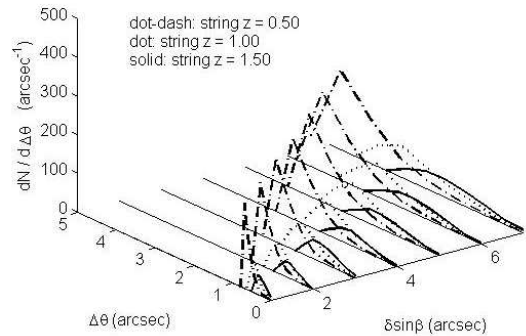}
\caption{\label{fig:sig}
Various examples of simulated signals where 
$dN/d\Delta\theta$ 
represents the number of lensed galaxy pairs per angular separation distance 
as a function 
of cosmic string parameters---energy-density ($\delta\sin\beta$) and redshift 
($z_l$).  
In these simulations, the number of signal events are scaled to a cosmic 
string with a total length of 10 arcminutes.}
\end{figure}

\protect
The simulation of galaxy image pairs caused by the presence of a cosmic 
string is accomplished by laying down sample strings of a chosen
energy-density/relative-tilt ($\delta\sin\beta$), 
and redshift ($z_l$) across our fiducial region.  
We then statistically tally the resulting galaxies that would have been 
``lensed'' if the string had existed.  
A galaxy is found to be lensed if the opening angle ($\Delta\theta$) 
calculated in Eq. \ref{eq:oa} is sufficient as to place the 
image-galaxy on the side of the string opposite the true-galaxy.  
The accumulated samples result in a catalog of lensed galaxies which also 
includes other pertinent information about each lensing event, such as the 
opening angle.  
This information is then used for computing signal densities as shown in 
Fig. \ref{fig:sig}.
To reduce numerical uncertainty, we simulate as many independent string trials 
as are necessary until the accumulated total of lensed sources ($N_{src}$) 
is 10,000.
That is, $1/\sqrt{N_{src}} < 0.01$ \cite{huterer, oguri}.  

\begin{figure}
\includegraphics{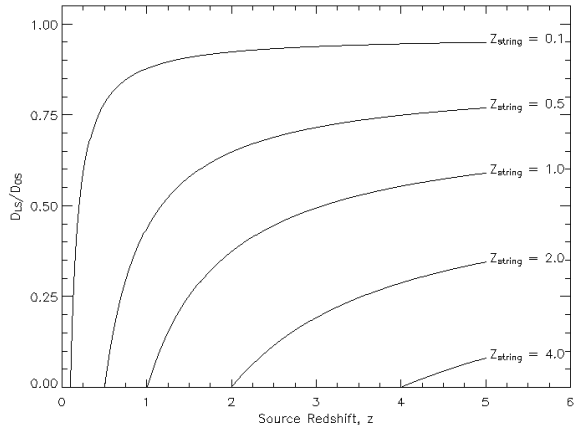}
\caption{\label{fig:ratio}
The ratio D$_{ls}/$D$_{os}$ is a multiplicative 
factor which affects the opening angle in Eq. \ref{eq:oa}.  
Here we plot it for a flat $\Lambda$CDM cosmology, 
$H_0$=73 km/(s Mpc), $\Omega_\Lambda=0.76$, $\Omega_M=0.24$.}
\end{figure}

The ratio D$_{ls}/$D$_{os}$ is a critical factor for determining opening 
angle, and hence, whether or not a lensed image results.  
We simulate these distance factors for a $\Lambda$CDM cosmology \cite{ryden}
with $H_0$ = 73 km/(s Mpc), $\Omega_\Lambda = 0.76$ and $\Omega_M = 0.24$.
The ratio of distance factors is illustrated in Fig. \ref{fig:ratio}.  
Knowledge of source redshift is essential in its calculation, however, 
redshift data for the GOODS survey is not presently available.  
This problem is resolved by assigning redshifts to each source 
based on their SExtractor I-band magnitude (MAG\_AUTO) as outlined 
in \cite{massey}.  
The source-number density is found to be reasonably approximated by:
\begin{equation}\label{eq:dNdz}
\frac{dN_{src}(z_s)}{dz} \propto z^{2}e^{-(z/z_m)^2}
\end{equation}
where
\begin{equation}\label{eq:zm}
z_m=0.722+0.149(I-22.0)
\end{equation}

\begin{figure}
\includegraphics{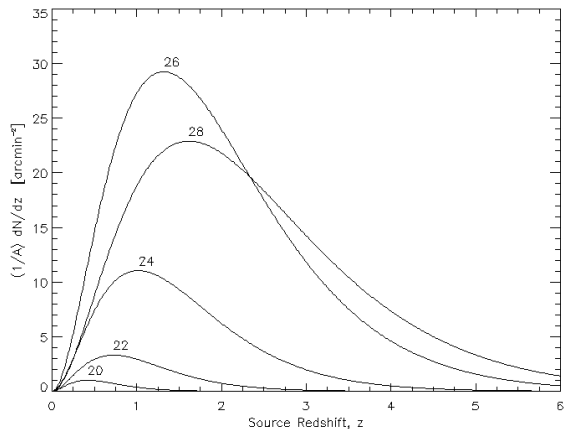}
\caption{\label{fig:pdf}
Modeled GOODS north field source redshift probability 
distribution function (PDF) as a function of redshift and I-band 
magnitude (MAG\_AUTO 
- shown on each curve).  Distributions are scaled to the north field area 
(A $\simeq 160$ arcmin$^2$).}
\end{figure}

Figure \ref{fig:pdf} is the result of applying Eq. \ref{eq:dNdz} to the 
GOODS north field catalog.  
Each source in the catalog is assigned a redshift that is distributed 
according to the model.  
The figure shows the number density of galaxies as a function of redshift in 
the GOODS north field for integer bins of I-band magnitude.

\section{\label{sec:eff}Detection Efficiencies:}\protect
There are a number of reasons why some lensed pairs would not be 
detected.
Figure \ref{fig:eff} shows a summary of the efficiencies as a function of the
opening angle between the galaxies and the string redshift.

\subsection{Survey edges}
\begin{figure}
\includegraphics{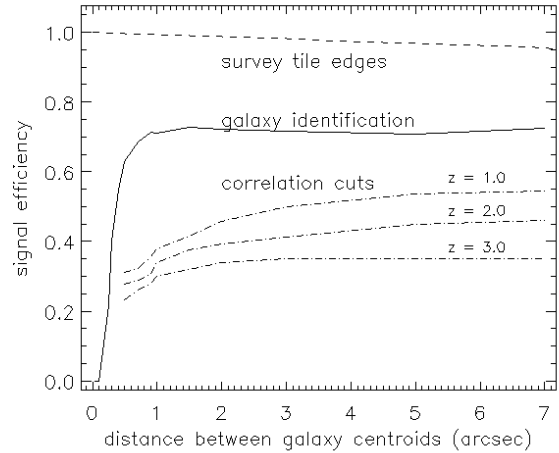}
\caption{\label{fig:eff}
Efficiency of detecting pairs of galaxies lensed by a cosmic string as a 
function of pair opening angle.  The correlation cut efficiencies are 
dependent on redshift as indicated.  The other efficiencies are only weakly
dependent on redshift.}
\end{figure}

Survey edges will occasionally be a factor when either the direct or lensed 
image of the galaxy does not land inside our fiducial region.  
We fit the shape of the background distibution in Fig. \ref{fig:pairs}
with a second order polynomial.
If the survey were infinitely large, the distribution would rise perfectly 
linearly as described in the overview, 
\begin{equation}\label{eq:linear}
dN/d\Delta\theta \propto 2\pi\delta\theta\Delta\theta
\end{equation}
where $\delta\theta$ is the bin width and $\Delta\theta$ is the opening
angle between the galaxies.
Because the measured distribution falls slightly below the line 
at the largest opening angles, we estimate the inefficiency 
due to the edges of our survey
as the ratio of the quadratic and linear terms in the $2^{nd}$ order fit to
the background.
As shown in Fig. \ref{fig:eff}, the inefficiency due to edges is never more 
than 5\% in our signal region.

\subsection{Galaxy pair detection and merging}
Given one galaxy, we estimate the probability that its lensed partner is 
found by embedding galaxies back into the original fits images. 
We use simulated strings to determine which galaxies are lensed in the
image.
So that the embedded galaxy does not contribute excess noise to the image, 
we smooth it with the 
Lee Filter \cite{LEEFILT} in IDL prior to adding its 
intensity to the image intensity.  
The smoothing makes the galaxy slightly dimmer than its original, and we note 
that this technique slightly underestimates the efficiency at large opening 
angles which results in our limits being slightly conservative. 
We then pass the image, including the embedded galaxies through the entire 
analysis chain.
 
The efficiencies are calculated by comparing the number of galaxies identified
by the analysis chain to
the number identified before embedding plus the number embedded.
We attribute the inefficiency at large opening angles to dim galaxies that
are lost due to measurement noise.  
At small opening angles SExtractor merges the galaxies into a single object.
The galaxy identification efficiency is only weakly dependent on the redshift 
of the string.

\subsection{Correlation cuts}
In addition to the inefficiencies inherent in finding the lensed galaxy,
there are inefficiencies due to the correlation estimators.
Noise can bias the $CORR$ and $XCORR$ variables. 
Using embedded galaxies that have been passed through the analysis chain, we 
compare the number that pass the correlation cuts to the number of embedded
galaxies that were detected.  
The efficiencies drop for small opening angles due to galaxies near the 
edge of our lensing corridor which are only partially lensed.  
The correlation efficiency also shows a relatively strong 
dependence on the redshift of the string.  
We attribute this to the fraction of dim galaxies lensed in the sample.
A high redshift string has a larger fraction of dim galaxies behind 
it than does a low redshift string.

\section{\label{sec:results}Results:}\protect
The distribution of matched galaxy pairs was shown in Fig. \ref{fig:pairs}.  
It rises nearly linearly as expected from Eq. \ref{eq:linear} with a
slight inefficiency due to the edges of the survey shown in 
Fig. \ref{fig:eff}.  For comparison, pairs from a cosmic string at 
a redshift of 1.2 and $\delta\sin\beta$ of $6\arcsec$ are included on 
the plot normalized to the mean string length of $13.1\arcmin$ in the 
north field and $11.7\arcmin$ in the south field.  
The upper curve is the simulated signal without detection inefficiencies
scaled from Fig. \ref{fig:sig}. 
The lower curve includes the measurement inefficiencies from 
Fig. \ref{fig:eff}.  

\begin{figure}
\includegraphics{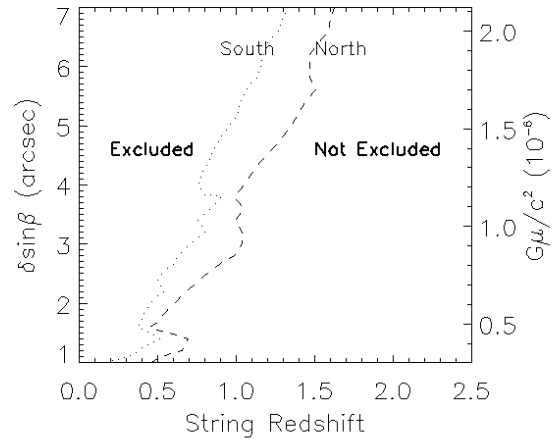}
\caption{\label{fig:limits}
Confidence limits at 95\% for lensed galaxies produced by a cosmic string as
a function of the string's mass and redshift.  The north field limits extend
to higher redshifts due to statistical fluctuations of signal and background, 
and also because the mean string length across the survey
is a bit longer and the number density of galaxies is somewhat higher.}
\end{figure}

We compare a wide variety of predicted cosmic string signals to the data to
determine limits.  For each signal, we integrate the signal to find $n_s$ 
pairs. We then integrate the matched pairs and the background curve over 
the opening angles from $0.4\arcsec$ where the efficiency estimates are 
reliable up to the point where there are no more signal pairs to
determine $n_{obs}$ and $n_b$.
We report classical single-sided Neyman confidence limits, (C.L.), where
the probability of finding a signal is limited to a region:
\begin{equation}\label{eq:cl}
P(n_{obs}<n_b+n_{lim}|\sigma) = C.L.
\end{equation} 
The estimator of the experimental fluctuations is $\sigma$, 
the mean background is $n_b$, and the minimum number of signal events
that would be consistent with background fluctuations is $n_{lim}$.  
That is, $n_s>n_{lim}$ is excluded and $n_s<n_{lim}$ is not.
Since our backgrounds are relatively large, we express the probability
as a Gaussian distribution with $\mu=(n_{obs}-n_b)$ and 
$\sigma = \sqrt{n_{obs}}$.
\begin{equation}\label{eq:prob}
P(n_x|\mu,\sigma) = \frac{1}{\sqrt{2\pi}\sigma}exp(-(\mu-n_x)^2/2\sigma^2)
\end{equation}
This is a Gaussian with mean near zero and a width that represents the 
statistical fluctuations in the data.  
The limit, $n_{lim}$, is then the value of $n_x$ for which 95\% of the area 
under the Gaussian is left of $n_x$.  
This is $P(n_x<n_{lim}|\mu,\sigma) = C.L.$.  
The resulting 95\% confidence limits are shown in Fig. \ref{fig:limits}.
The limits extend from $0.5\arcsec<\delta\sin\beta<7\arcsec$.  
Taking the mean tilt of a string with respect to the observer to be 
$<sin\beta> = 2/\pi$ we relate the opening angle to the mass scale by
$G\mu/c^2 = \delta<\sin\beta>/(8\pi)$ shown on the right-hand axis.
We see no evidence for cosmic strings out to a redshift greater than 
0.5 in the north field and greater than 0.3 in the south field 
and place a limit on $G\mu/c^2<3.0\times10^{-7}$ at 95\% C.L.
The north field limits extend to higher redshifts than
the south field limits due to statistical fluctuations of signal and background
and also because the north field has a longer mean string 
length across the survey and a higher number density of galaxies.

\begin{figure}
\includegraphics{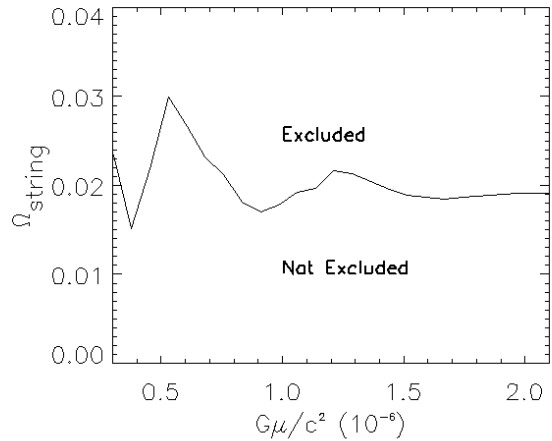}
\caption{\label{fig:omega}
Confidence limits at 95\% on $\Omega_{String}$ as a function of the string
mass.}
\end{figure}

It is highly unlikely that a string passes through both the north and south 
fields and we therefore treat the two surveys as uncorrelated measures of the
string density.  
If strings are rare occurrences, it is possible that none would appear in the 
GOODS fields and that other survey fields may yield different results.  
We can interpret our result, however, as excluding the possibility
that 3 strings would be located in any one field of view with 95\% confidence.
This corresponds as a string density
\begin{equation}\label{eq:omega}
\Omega_{strings} = \frac{\rho_{strings}}{\rho_{critical}}
\simeq \frac{N\mu}{fov\  \eta^2/3}\times\frac{8\pi G}{3H_0^2}
\end{equation}
where $N=3$ strings, $\mu$ is the mass per length of the string, $fov$ is the
survey field of view (159.5 arcmin$^2$ in the north and 155.4 arcmin$^2$ in the
south), and $\eta$ is the comoving distance.  
We combine the
north and south fields by averaging the string length within the field of view
and summing the volume of the two fields.
Figure \ref{fig:omega} shows the string densities excluded by this method.
The limit excludes a string density that is 2\% of the critical density.
The variation observed is due to statistical fluctuations in background 
and is sensitive to the small-angle signal efficiencies.

\section{\label{sec:conclusion}Conclusion:}\protect
We use the GOODS HDF-N and CDF-S fields to search for cosmic strings.  
We find no evidence for the existence of the gravitational lensing 
signature.  
We have included the observational efficiencies in our analysis using a 
new technique of embedding galaxies based on a robust string simulation.  
Our results are summarized in Figs. \ref{fig:limits} and \ref{fig:omega}.
From this search we conclude with 95\% confidence that
$G\mu/c^2<3.0\times10^{-7}$ out to redshifts greater than 0.5 and 
that $\Omega_{strings}<0.02$.  We note that these results are for 
long straight strings, but also exclude moderately curved strings as long as there
are no kinks in the field of view.

These are the first reported limits using the gravitational lensing 
signature and are consistent with recently reported sensitivities of such 
searches \cite{gasparini}.
Our excluded masses are smaller than those excluded by direct
CMB searches \cite{jeong, lo}.
Our limits on $\Omega_{string}$ are still weaker, however, due to the 
full-sky coverage and high redshift of the CMB.  
Our excluded masses are larger than those reported by 
parameter fits to the CMB \cite{WMAP-3}
as well as gravitational wave searches \cite{pulsar, ligo},
but these depend on features of the modeling which do not affect our 
direct search.  
We note that recently rekindled interest in microlensing
signatures \cite{micro-1, micro-2} is motivated by the challenge of 
detecting smaller mass cosmic strings over a larger fraction of the sky out
to high redshift.  We are now in the infancy of wide-field high 
resolution surveys and are excited about the prospect of pursuing these
techniques with larger surveys such as COSMOS\cite{cosmos} and SNAP\cite{snap}.

\section{\label{sec:ack}Acknowledgements:}\protect
We would like to thank Lluvia Zuniga, Kirsten Howley, Wes Salpiea, 
and Mia Ihm for their early involvement in this project.  
E. Albin was partially supported by the SULI summer program at LBNL.
K. James was partially supported by the Cal Poly SLO Department of Physics.
J. L. Christiansen and E. Albin were partially supported by the  
Cal Poly SLO Provost's Extramural Funding Initiative.

\bibliography{../strings/string.bib}

\begin{thebibliography}{29}
\expandafter\ifx\csname natexlab\endcsname\relax\def\natexlab#1{#1}\fi
\expandafter\ifx\csname bibnamefont\endcsname\relax
  \def\bibnamefont#1{#1}\fi
\expandafter\ifx\csname bibfnamefont\endcsname\relax
  \def\bibfnamefont#1{#1}\fi
\expandafter\ifx\csname citenamefont\endcsname\relax
  \def\citenamefont#1{#1}\fi
\expandafter\ifx\csname url\endcsname\relax
  \def\url#1{\texttt{#1}}\fi
\expandafter\ifx\csname urlprefix\endcsname\relax\def\urlprefix{URL }\fi
\providecommand{\bibinfo}[2]{#2}
\providecommand{\eprint}[2][]{\url{#2}}

\bibitem[{\citenamefont{Kibble}(1976)}]{kibble-GUT}
\bibinfo{author}{\bibfnamefont{T.}~\bibnamefont{Kibble}}, \bibinfo{journal}{J.
  Phys. A} \textbf{\bibinfo{volume}{9}}, \bibinfo{pages}{1387}
  (\bibinfo{year}{1976}).

\bibitem[{\citenamefont{Polchinski}()}]{polchinski-DF}
\bibinfo{author}{\bibfnamefont{J.}~\bibnamefont{Polchinski}},
  \eprint{hep-th/0412244}.

\bibitem[{\citenamefont{Hindmarsh and Kibble}(1995)}]{hindmarsh}
\bibinfo{author}{\bibfnamefont{M.}~\bibnamefont{Hindmarsh}} \bibnamefont{and}
  \bibinfo{author}{\bibfnamefont{T.}~\bibnamefont{Kibble}},
  \bibinfo{journal}{Rep. Prog. Phys.} \textbf{\bibinfo{volume}{58}},
  \bibinfo{pages}{477} (\bibinfo{year}{1995}).

\bibitem[{\citenamefont{Sakellariadou}()}]{sakellariadou}
\bibinfo{author}{\bibfnamefont{M.}~\bibnamefont{Sakellariadou}},
  \eprint{hep-th/0602276}.

\bibitem[{\citenamefont{Pogosian et~al.}(2003)\citenamefont{Pogosian, Tye,
  Wasserman, and Wyman}}]{WMAP-1}
\bibinfo{author}{\bibfnamefont{L.}~\bibnamefont{Pogosian}},
  \bibinfo{author}{\bibfnamefont{S.}~\bibnamefont{Tye}},
  \bibinfo{author}{\bibfnamefont{I.}~\bibnamefont{Wasserman}},
  \bibnamefont{and} \bibinfo{author}{\bibfnamefont{M.}~\bibnamefont{Wyman}},
  \bibinfo{journal}{Phys. Rev. D} \textbf{\bibinfo{volume}{68}},
  \bibinfo{pages}{023506} (\bibinfo{year}{2003}).

\bibitem[{\citenamefont{Pogosian et~al.}(2004)\citenamefont{Pogosian, Wyman,
  and Wasserman}}]{WMAP-2}
\bibinfo{author}{\bibfnamefont{L.}~\bibnamefont{Pogosian}},
  \bibinfo{author}{\bibfnamefont{M.}~\bibnamefont{Wyman}}, \bibnamefont{and}
  \bibinfo{author}{\bibfnamefont{I.}~\bibnamefont{Wasserman}},
  \bibinfo{journal}{J. Cosmol. Astropart.} \textbf{\bibinfo{volume}{09}},
  \bibinfo{pages}{008} (\bibinfo{year}{2004}).

\bibitem[{\citenamefont{Wyman et~al.}(2006)\citenamefont{Wyman, Pogosian, and
  Wasserman}}]{WMAP-3}
\bibinfo{author}{\bibfnamefont{M.}~\bibnamefont{Wyman}},
  \bibinfo{author}{\bibfnamefont{L.}~\bibnamefont{Pogosian}}, \bibnamefont{and}
  \bibinfo{author}{\bibfnamefont{I.}~\bibnamefont{Wasserman}},
  \bibinfo{journal}{Phys. Rev. D} \textbf{\bibinfo{volume}{73}},
  \bibinfo{pages}{089905} (\bibinfo{year}{2006}).

\bibitem[{\citenamefont{Bevis et~al.}(2007)\citenamefont{Bevis, Hindmarsh,
  Kunz, and Urrestilla}}]{WMAP-4}
\bibinfo{author}{\bibfnamefont{N.}~\bibnamefont{Bevis}},
  \bibinfo{author}{\bibfnamefont{M.}~\bibnamefont{Hindmarsh}},
  \bibinfo{author}{\bibfnamefont{M.}~\bibnamefont{Kunz}}, \bibnamefont{and}
  \bibinfo{author}{\bibfnamefont{J.}~\bibnamefont{Urrestilla}},
  \bibinfo{journal}{Phys. Rev. D} \textbf{\bibinfo{volume}{75}},
  \bibinfo{pages}{065015} (\bibinfo{year}{2007}).

\bibitem[{\citenamefont{Jeong and Smoot}(2005)}]{jeong}
\bibinfo{author}{\bibfnamefont{E.}~\bibnamefont{Jeong}} \bibnamefont{and}
  \bibinfo{author}{\bibfnamefont{G.}~\bibnamefont{Smoot}},
  \bibinfo{journal}{Astrophys. J.} \textbf{\bibinfo{volume}{624}},
  \bibinfo{pages}{21} (\bibinfo{year}{2005}).

\bibitem[{\citenamefont{Lo and Wright}()}]{lo}
\bibinfo{author}{\bibfnamefont{A.}~\bibnamefont{Lo}} \bibnamefont{and}
  \bibinfo{author}{\bibfnamefont{E.}~\bibnamefont{Wright}},
  \eprint{astro-ph/0503120}.

\bibitem[{\citenamefont{Jenet et~al.}(2006)}]{pulsar}
\bibinfo{author}{\bibfnamefont{F.}~\bibnamefont{Jenet}} \bibnamefont{et~al.},
  \bibinfo{journal}{Astrophys. J.} \textbf{\bibinfo{volume}{653}},
  \bibinfo{pages}{1571} (\bibinfo{year}{2006}).

\bibitem[{\citenamefont{Siemens et~al.}(2007)\citenamefont{Siemens, Mandic, and
  Creighton}}]{ligo}
\bibinfo{author}{\bibfnamefont{X.}~\bibnamefont{Siemens}},
  \bibinfo{author}{\bibfnamefont{V.}~\bibnamefont{Mandic}}, \bibnamefont{and}
  \bibinfo{author}{\bibfnamefont{J.}~\bibnamefont{Creighton}},
  \bibinfo{journal}{Phys. Rev. Lett.} \textbf{\bibinfo{volume}{98}},
  \bibinfo{pages}{111101} (\bibinfo{year}{2007}).

\bibitem[{\citenamefont{Sazhin et~al.}(2003)}]{csl-1}
\bibinfo{author}{\bibfnamefont{M.}~\bibnamefont{Sazhin}} \bibnamefont{et~al.},
  \bibinfo{journal}{Mon. Not. Roy. Astron. Soc.}
  \textbf{\bibinfo{volume}{343}}, \bibinfo{pages}{353} (\bibinfo{year}{2003}).

\bibitem[{\citenamefont{Sazhin et~al.}()}]{csl-2}
\bibinfo{author}{\bibfnamefont{M.}~\bibnamefont{Sazhin}} \bibnamefont{et~al.},
  \eprint{astro-ph/0406516}.

\bibitem[{\citenamefont{Agol et~al.}(2006)\citenamefont{Agol, Hogan, and
  Plotkin}}]{csl-1-refuted}
\bibinfo{author}{\bibfnamefont{E.}~\bibnamefont{Agol}},
  \bibinfo{author}{\bibfnamefont{C.}~\bibnamefont{Hogan}}, \bibnamefont{and}
  \bibinfo{author}{\bibfnamefont{R.}~\bibnamefont{Plotkin}},
  \bibinfo{journal}{Phys. Rev. D} \textbf{\bibinfo{volume}{73}},
  \bibinfo{pages}{087302} (\bibinfo{year}{2006}).

\bibitem[{\citenamefont{Oguri and Takahashi}(2005)}]{oguri}
\bibinfo{author}{\bibfnamefont{M.}~\bibnamefont{Oguri}} \bibnamefont{and}
  \bibinfo{author}{\bibfnamefont{K.}~\bibnamefont{Takahashi}},
  \bibinfo{journal}{Phys. Rev. D} \textbf{\bibinfo{volume}{72}},
  \bibinfo{pages}{085013} (\bibinfo{year}{2005}).

\bibitem[{\citenamefont{Gasparini et~al.}()}]{gasparini}
\bibinfo{author}{\bibfnamefont{M.}~\bibnamefont{Gasparini}}
  \bibnamefont{et~al.}, \eprint{astro-ph/0710.5544}.

\bibitem[{\citenamefont{Peebles}(1993)}]{peebles}
\bibinfo{author}{\bibfnamefont{P.}~\bibnamefont{Peebles}},
  \emph{\bibinfo{title}{Principles of Physical Cosmology}}
  (\bibinfo{publisher}{Princeton University Press}, \bibinfo{year}{1993}).

\bibitem[{\citenamefont{Ryden}(2003)}]{ryden}
\bibinfo{author}{\bibfnamefont{B.}~\bibnamefont{Ryden}},
  \emph{\bibinfo{title}{Introduction to Cosmology}}
  (\bibinfo{publisher}{Pearson Education Inc.}, \bibinfo{year}{2003}).

\bibitem[{\citenamefont{Giavalisco and others (GOODS~Team)}(2004)}]{GOODS}
\bibinfo{author}{\bibfnamefont{M.}~\bibnamefont{Giavalisco}} \bibnamefont{and}
  \bibinfo{author}{\bibnamefont{others (GOODS~Team)}},
  \bibinfo{journal}{Astrophys. J.} \textbf{\bibinfo{volume}{L93}},
  \bibinfo{pages}{600} (\bibinfo{year}{2004}).

\bibitem[{\citenamefont{Bertin}(2005)}]{SEX-1}
\bibinfo{author}{\bibfnamefont{E.}~\bibnamefont{Bertin}},
  \emph{\bibinfo{title}{Sextractor v2.4 user's manual}} (\bibinfo{year}{2005}).

\bibitem[{\citenamefont{Bertin and Arnouts}(1996)}]{SEX}
\bibinfo{author}{\bibfnamefont{E.}~\bibnamefont{Bertin}} \bibnamefont{and}
  \bibinfo{author}{\bibnamefont{Arnouts}}, \bibinfo{journal}{Astron. and
  Astron. Suppl. Ser.} \textbf{\bibinfo{volume}{117}}, \bibinfo{pages}{393}
  (\bibinfo{year}{1996}).

\bibitem[{\citenamefont{Lee}(1986)}]{LEEFILT}
\bibinfo{author}{\bibfnamefont{J.}~\bibnamefont{Lee}}, \bibinfo{journal}{Opt.
  Eng.} \textbf{\bibinfo{volume}{25}}, \bibinfo{pages}{636}
  (\bibinfo{year}{1986}).

\bibitem[{\citenamefont{Huterer and Vachaspati}(2003)}]{huterer}
\bibinfo{author}{\bibfnamefont{D.}~\bibnamefont{Huterer}} \bibnamefont{and}
  \bibinfo{author}{\bibfnamefont{T.}~\bibnamefont{Vachaspati}},
  \bibinfo{journal}{Phys. Rev. D} \textbf{\bibinfo{volume}{68}},
  \bibinfo{pages}{041301} (\bibinfo{year}{2003}).

\bibitem[{\citenamefont{Massey et~al.}(2004)}]{massey}
\bibinfo{author}{\bibfnamefont{R.}~\bibnamefont{Massey}} \bibnamefont{et~al.},
  \bibinfo{journal}{Astron.J.} \textbf{\bibinfo{volume}{127}},
  \bibinfo{pages}{3089} (\bibinfo{year}{2004}).

\bibitem[{\citenamefont{Mack et~al.}(2007)\citenamefont{Mack, Wesley, and
  King}}]{micro-1}
\bibinfo{author}{\bibfnamefont{K.}~\bibnamefont{Mack}},
  \bibinfo{author}{\bibfnamefont{H.}~\bibnamefont{Wesley}}, \bibnamefont{and}
  \bibinfo{author}{\bibfnamefont{L.}~\bibnamefont{King}},
  \bibinfo{journal}{Phys. Ref. D} \textbf{\bibinfo{volume}{76}},
  \bibinfo{pages}{123515} (\bibinfo{year}{2007}).

\bibitem[{\citenamefont{Kuijken et~al.}()\citenamefont{Kuijken, Siemens, and
  Vachaspati}}]{micro-2}
\bibinfo{author}{\bibfnamefont{K.}~\bibnamefont{Kuijken}},
  \bibinfo{author}{\bibfnamefont{X.}~\bibnamefont{Siemens}}, \bibnamefont{and}
  \bibinfo{author}{\bibfnamefont{T.}~\bibnamefont{Vachaspati}},
  \eprint{astro-ph/0707.2971}.

\bibitem[{\citenamefont{Scoville et~al.}()}]{cosmos}
\bibinfo{author}{\bibfnamefont{N.}~\bibnamefont{Scoville}}
  \bibnamefont{et~al.}, \eprint{astro-ph/0612305}.

\bibitem[{\citenamefont{Aldering et~al.}()}]{snap}
\bibinfo{author}{\bibfnamefont{G.}~\bibnamefont{Aldering}}
  \bibnamefont{et~al.}, \eprint{astro-ph/0405232}.

\end{thebibliography}

\end{document}